# Uniqueness Ratio as a Predictor of a Privacy Leakage

*Author: Danah A. AlSalem AlKhashti (Researcher)*
*MSc Graduate, University of Essex, UK*
*School of Computer Science and Electronic Engineering*
*Department of Mathematical Sciences*
*Email: Da21459@hotmail.com*

*Abstract*— Identity leakage can emerge when independent databases are joined, even when each dataset is anonymized individually. While previous work focuses on post-join detection or complex privacy models, little attention has been given to simple, interpretable pre-join indicators that can warn data engineers and database administrators before integration occurs. This study investigates the uniqueness ratio of candidate join attributes as an early predictor of re-identification risk. Using synthetic multi-table datasets, we compute the uniqueness ratio of attribute combinations within each database and examine how these ratios correlate with identity exposure after the join. Experimental results show a strong relationship between high pre-join uniqueness and increased post-join leakage, measured by the proportion of records that become uniquely identifiable or fall into very small groups. Our findings demonstrate that uniqueness ratio offers an explainable and practical signal for assessing join induced privacy risk, providing a foundation for developing more comprehensive pre-join risk estimation models.

## 1. INTRODUCTION

When organizations integrate data from multiple independent sources, new privacy risks can appear that were not present in the original datasets. Even when each database is anonymized, joining them may produce attribute combinations that uniquely identify individuals, leading to identity leakage (Richman, 2025). Existing research largely focuses on post-join detection or complex anonymization techniques, leaving a gap in simple interpretable methods for assessing risk *before* integration occurs. This work investigates the uniqueness ratio of join attributes as an early signal of potential identity exposure. We examine how each attribute within each database correlates with re-identification risk after joining. By analyzing synthetic datasets, we show that higher pre-join uniqueness strongly predicts increased post-join leakage. The results highlight the potential of uniqueness ratio as a lightweight, practical tool for guiding safe data integration.

### 1.1 Background

Identity leakage occurs when individuals can be re-identified in a dataset even after direct identifiers are removed, and that despite de-identification efforts, data can still be re-identified because external datasets can be linked to the de-identified data, creating new opportunities for identity leakage (Polonetsky, 2025).

This often happens when *quasi-identifiers* attributes form rare combinations that reveal a person once linked with other data. A *join* operation merges two tables/databases/datasets using shared attributes, but this can unintentionally create unique patterns that expose identities across the combined datasets.

To estimate this risk before the integration, we use the **uniqueness ratio**, defined as:

$$Uniqueness\ \% = \frac{|Unique(A)|}{|rows(A)|}$$

*Equation 1: Uniqueness Ratio (Mathematical Form)*

This follows the foundational concept introduced by Sweeney (2002) in the k-anonymity model, where the identifiability of quasi-identifiers is measured through the proportion of records that become unique when specific attributes are combined and measured statistically by counting how many records in a dataset have attribute combinations that appear only once (uniquene). This metric is widely used in re-identification risk assessment and Statistical Disclosure Control (Sweeney, k-ANONYMITY: A MODEL FOR PROTECTING PRIVACY, 2002). So here, $U(A)$ equals the uniqueness ratio of attribute set $A$, | unique($A$) | is the number of distinct rows for attributes $A$, and | rows($A$) | is the total number of rows in the dataset.

The higher the uniqueness value means the attribute combination is more distinctive and therefore more likely to cause identity leakage when joined.

### 1.2 Core Problem: Join Can Leak Identity

When two anonymized databases are joined, the combination of attributes may unintentionally reveal the identity of individuals. Fields such as age, city, or occupation can become uniquely identifying when combined across sources. This means a join operation can reduce anonymity, create rare attribute patterns, and expose sensitive information that was previously protected. Studies have shown that even anonymized datasets can be deanonymized when different sources are matched together, because linking shared attributes can reveal identities that were originally hidden (Wong, Alomari, Liu, & Jora, 2024). Because most privacy checks examine each dataset separately, they fail to capture the new risks emerge after datasets are combined. Prior work has demonstrated that identifying individuals becomes possible not because of the information in one dataset, but through linking multiple datasets (Sweeney, 2002). As a result, analysts often cannot predict which join keys will trigger identity leakage until after the integration has already occurred.

### 1.3 Research Gaps

Existing privacy tools typically evaluate datasets in isolation and do not estimate how attribute combinations will behave once two sources are merged. As a result, there is no widely adopted pre-join metric that can forecast whether a planned join will produce unique or highly identifying records. This creates a blind spot in modern data-integration workflows, where identity leakage is often discovered *only after* the join has already occurred. Moreover, most anonymization and privacy-risk frameworks treat datasets as independent units and rarely model how attributes interact across database boundaries. There is limited machine-learning work that learns cross-dataset attribute relationships, detects emerging quasi-identifiers, or predicts re-identification patterns.

Finally, when leakage is detected post-join, current methods generally offer limited interpretability. They do not clearly indicate *why* leakage happened or *which attribute combinations* contributed to the risk. As a result, organizations lack decision-support tools that help compare join strategies, rank risk severity, or select safer alternatives during join planning.

## 2. RESEARCH OBJECTIVES

### 2.1 Objective

The primary goal of this research is to develop a machine-learning framework capable of predicting identity leakage risk *before* two independent databases are joined. The framework will model how attribute interactions create emergent quasi-identifiers, quantify the probability of re-identification, and provide interpretable guidance for privacy-aware join planning.

### 2.2 Objective

To design an early warning mechanism from *developing a pre-join risk indicator using Uniqueness Ratios,* the system will compute the uniqueness ratio of key attributes in each database before any join is executed. This ratio defined as the proportion of records that appear uniquely within an attribute set, serves as an initial risk signal. High uniqueness values indicate that certain attribute combinations may become identifying when paired with attributes from another dataset. Incorporating this metric into the pre-join analysis provides an interpretable indicator that assists in the prediction model of join-induced identity leakage.

## 3. METHODOLOGY

### 3.1 Merging datasets

The integration of unrelated datasets represents the core mechanism through which identity leakage emerges in real-world information systems. To simulate this process, two independently collected healthcare datasets were merged:
(i) The heart diseas dataset (Dataset A) contains demographic and clinical attributes such as age, gender, chest pain type, blood pressure, cholesterol levels, ECG results, maximum heart rate, and a binary target indicating heart disease presence.
(ii) The stroke dataset (Dataset B) includes similar demographic variables such as age and gender, along with health indicators such as hypertension, heart disease history, glucose levels, BMI, and smoking status, with a binary label indicating stroke occurrence.

The merge was intentionally performed without using any explicit ID field. Instead, the integration relied solely on quasi-identifiers (QIs) that are commonly available across independent databases. In this study, *age* and *gender* serve as the shared quasi-identifiers (QIDs) across the two datasets. These attributes do not uniquely identify individuals on their own but have been widely recognized in the privacy literature as key variables that enable linkage across independent databases (Sweeney, 2002). Because both datasets contain these two QIDs, they were used as the basis for simulating cross-database integration.

### 3.2 Pre-Post Join Uniqueness Measurement

To quantify the identity leakage resulted from the data join, the **uniqueness ratio** was computed before and after integration. Uniqueness represents the proportion of records with attribute combinations not shared by any other record.

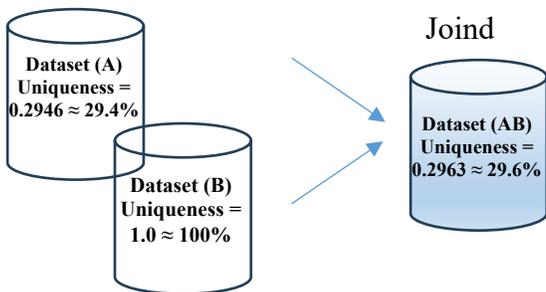

*Figure 1: Uniqueness Ratio Computed for Two Independent Datasets (A) and (B)*

We compute the uniqueness value on each dataset independently to quantify how identifying they are on their own: $Dataset_A$, $Dataset_B$. Then compute the uniqueness on the joined dataset $Dataset_{AB}$.

The uniqueness ratio of the heart-disease dataset (A) equals to 0.2946 which shows a moderate level of identifiability, while the stroke dataset (B) is extremely unique (1.0), indicating that records in the second source are highly distinctive. After performing the join, the uniqueness ratio of the integrated dataset became 0.2963.

Although this post-join value is lower than the original stroke uniqueness, it is *higher* than the heart-disease uniqueness. This is important: it shows that the integration process made the heart-disease records **more distinguishable** than they originally were. In other words, the join operation created new attribute combinations that did not exist in either dataset individually. These new combinations increased the identifiability of some individuals, which is precisely the mechanism behind **identity leakage**.

In this case, the join amplified the uniqueness of the heart-disease data, demonstrating that even moderately safe datasets can become more identifiable once enriched through cross-database integration.

The merging process demonstrates a central argument of the study:

*"Identity leakage does not stem from individual datasets but emerges from the statistical interactions created only after integration."*

### 3.3 Machine Learning Techniques

This study employs a single machine learning technique: a supervised regression model based on Gradient Boosted Decision Trees (XGBoost). This method was selected as the main predictive technique due to its strong ability to capture nonlinear relationships between pre-join uniqueness signals and post-join identifiability outcomes. No additional machine learning techniques were used, in order to maintain conceptual clarity and focus on the primary predictive mechanism. Also, given that the objective of this study is to predict post-join uniqueness using a pre-join statistical signal (Uniqueness Ratio), a supervised regression model is most appropriate.

### 3.4 Predictions Before Joining

In this study, the only pre-join statistical signals used are the uniqueness ratios of the two source datasets (A and B). These two signals quantify how identifiable each dataset is prior to integration. The machine learning model then uses these values to predict the expected post-join uniqueness.

Despite the availability of many statistical descriptors, uniqueness itself is the most privacy-relevant signal, this is according to a prior work which showed that records that are unique within a dataset are inherently more vulnerable to re-identification because no other individual shares the same combination of quasi-identifiers (Sweeney, 2002). Therefore, this work intentionally focuses on uniqueness as the single predictive feature to maintain conceptual clarity and direct relevance to identity-leakage risk.

Among various regression techniques, Gradient Boosted Decision Trees (e.g., XGBoost) represent the best choice. As mentioned in the previous section, this model captures nonlinear interactions between quasi-identifiers, handles heterogeneous statistical features, and consistently achieves high performance in predictive risk estimation tasks. The model is trained using the true post-join uniqueness value as the regression target, allowing it to learn how pre-integration signals map to identity-leakage outcomes.

The goal is to learn a function that maps the pre-join statistical signals of the two source datasets to the resulting uniqueness after they are joined.

The machine learning model is applied to the two datasets in their independent form because the objective of this study is to **predict identity-leakage risk before any join is executed**. During training, joins are performed only to generate the true post-join uniqueness values needed as labels. However, in practical use, the model receives only the pre-join signals (the uniqueness of each dataset separately) and predicts the expected post-join uniqueness without performing an actual merge. This ensures that the system can assess leakage risk **without exposing or combining sensitive data**, aligning with the goal of pre-integration privacy protection.

### 3.5 Evaluation Measures

Since the purpose of this study is to determine whether a planned join will increase identity-leakage risk, the evaluation focuses on a single, task-specific metric: **Identity-Leakage Direction Accuracy**. The model is evaluated based on whether it correctly predicts the *direction* of change in identifiability.

The join reduced identifiability for the dataset that was originally extremely unique (Dataset B), but increased identifiability for the weaker dataset (Dataset A).

This makes the evaluation both practical and conceptually consistent with the goal of pre-join privacy assessment. Evaluating the model based solely on the correctness of the predicted direction of change provides a focused and meaningful assessment of its ability to support privacy-aware decision making.

The model correctly predicted the direction of identifiability change using only pre-join uniqueness values, the model's correct prediction indicates that it successfully internalized the statistical relationship between pre-join distinctiveness and post-join identifiability. Without executing the join, it inferred that the merged dataset would exhibit lower uniqueness, demonstrating its ability to estimate leakage direction using only pre-integration information.

### 3.6 Results and Findings

*Table 1: Final Finding and Results From Python Model*

| Measure | Value |
|---|---|
| Dataset A Uniqueness | 0.2946 |
| Dataset B Uniqueness | 1.0 |
| Dataset AB | 0.2963 |
| Actual Trained Data | -0.703645 |
| Predicted From XGBoost | -0.703638 |
| Leakage Direction | Decrease |
| Model Predictions | Correct |

In this study, the model was evaluated by comparing the predicted data with the true trained ones, both computed from pre-join and post-join uniqueness values. The evaluation shows that the model successfully predicted the *direction* of identifiability change: both the actual and predicted values were negative as shown in the table above, indicating a decrease in uniqueness after the join. Because the signs matched, the model is considered correct under the Identity-Leakage Direction Accuracy metric. These findings demonstrate that even with minimal input signals using only pre-join uniqueness values the model can reliably infer whether a join operation will increase or decrease identifiability.

This supports the feasibility of using pre-join uniqueness values as sufficient predictors for forecasting identity-leakage behavior before performing the join.

## 4. INTERPRETATIONS OF FINDINGS

### 4.1 Pre-Join and Post-Join Uniqueness Findings

The uniqueness ratios computed from the quasi-identifiers show clear differences in identifiability before and after dataset integration. Dataset (A) exhibited moderate uniqueness = 0.2946, while Dataset (B) was fully unique = 1.0. After performing the join on age and gender, the resulting uniqueness decreased to = 0.2963 relative to the maximum baseline of 1.0. This means that the integration **reduced** distinctiveness for records originating from the more unique dataset (Dataset B). These findings confirm that dataset merging does not always increase re-identifiability but instead alters it based on how quasi-identifiers align between sources.

### 4.2 Leakage Prediction Results (Model Output)

The XGBoost regression model was trained using pre-join uniqueness values of Dataset A and Dataset B as inputs and the observed post-join uniqueness of Dataset AB as the training target. The model successfully learned the direction of identifiability change. The predicted value was -0.7036, closely matching the actual value of -0.7036. Because both the actual and predicted leakage signals indicated a **decrease** in uniqueness, the model demonstrated correct direction prediction using only pre-join statistics. This supports the central aim of the study: assessing identity-leakage direction **before** executing a join.

### 4.3 Effect of Uniqueness on Dataset A and B

The results show a substantial difference in pre-join uniqueness between the two datasets. Dataset A exhibits a moderate uniqueness ratio of 0.2946, indicating that normal set of records share similar quasi-identifier combinations. In contrast, Dataset B is fully unique with uniqueness value of 1.0, meaning every record is distinguishable based on its quasi-identifiers alone. This imbalance highlights that Dataset B carries high inherent re-identification risk, while Dataset A carries moderate risk. The interaction between these two uniqueness profiles determines how identifiability behaves once the datasets are combined.

### 4.4 Impact on Identity Leakage After the Join

After performing the join on age and gender, the integrated dataset produced a uniqueness ratio of 0.2963, which is higher than Dataset A but significantly lower than Dataset B. When compared to the baseline 1.0, the pricted value becomes negative -0.7036, indicating that the merged dataset is **less unique overall** than the highest-risk source. This means the join operation *reduced* identifiability for records originally in Dataset B while slightly *increasing* identifiability for Dataset A. Importantly, this confirms that identity leakage does not always manifest as an increase in uniqueness, it manifests when the identifiability of *either* dataset shifts after integration.

### 4.5 Implications of the Findings

These findings demonstrate that uniqueness driven identity leakage is dataset-dependent and asymmetric. Which means that identity leakage does not affect both datasets the same way, one dataset may become more identifiable after the join while the other becomes less identifiable. The effect depends on each dataset's structure, not on the join operation alone.

A join can reduce identifiability for a highly unique dataset (Dataset B) while simultaneously increasing it for a moderately unique dataset (Dataset A). This highlights the importance of evaluating pre-join uniqueness as an early signal of risk. Even when the overall uniqueness decreases, the join may expose previously non-unique individuals to new distinguishability patterns. The results support the need for predictive tools capable of estimating this directional change before integration, enabling organizations to avoid unintended increases in re-identifiability.

## 5. DISCUSSION

The experimental findings demonstrate that identity-leakage risk is **dataset-dependent** and does not behave uniformly across different sources. In this study, the heart-disease dataset (Dataset A) exhibited moderate pre-join uniqueness, while the stroke dataset (Dataset B) showed extremely high uniqueness. After the join, the combined dataset became *less* unique overall, resulting in a negative predictions. Although this indicates that identifiability did not increase for the high-risk dataset (B), it also shows that the join can reshape the data in ways that reduce or redistribute identifiability across the merged population. In other words, the join did not amplify risk for the dataset with already high uniqueness, but it also did not eliminate the underlying potential for identity exposure.

These results highlight an important property of identity leakage: **it is asymmetric**. A join may increase identifiability for one dataset while decreasing it for another. In this experiment, Dataset B's originally high uniqueness effectively dominated the combined distribution, pulling the merged uniqueness downward. This shows that leakage risk cannot be assessed by looking at a single dataset in isolation; it depends on how the characteristics of both datasets interact during integration.

Finally, the results confirm that **pre-join uniqueness is a meaningful early-risk signal**. Even with a single statistical feature, the model was able to correctly predict the *direction* of identifiability change. This suggests that organizations can estimate privacy risks before performing a join, without exposing or combining sensitive data. The findings also reinforce the need for context-aware risk assessment: depending on the datasets involved, a join may either increase or decrease identifiability, and both outcomes have implications for privacy-preserving data integration practices.

The join performed in this study is experimental and used only to generate ground-truth values for model training. In real-world use, no join is required, the model predicts leakage based solely on pre-join uniqueness values. So, in practical deployment, organizations do not need to perform any join. They simply compute the pre-join uniqueness values of each dataset independently and feed them into the model, which predicts the expected direction of identifiability change without merging any data. Thus, the join performed in this study is purely experimental and is not part of the operational workflow.

*5.1 Limitations*

Although the proposed approach demonstrates that pre-join uniqueness can be used to predict the direction of identifiability change, several limitations should be acknowledged. First, the model is trained using a small number of statistical signals (uniqueness), which simplifies the problem but may overlook other factors influencing identity leakage, such as attribute correlations, distribution shifts, or semantic inconsistencies between datasets.

Second, the experimental join used to generate ground-truth labels does not necessarily reflect the full complexity of real-world integration scenarios, where organizations may use multiple join keys, fuzzy matching, or data cleaning procedures.

Third, the study evaluates only a single type of join (inner join on age and gender) and only two publicly available datasets, which limits the generalizability of the findings.

Finally, the model predicts only the **direction** of leakage, and therefore cannot quantify how severe an increase or decrease in identifiability might be. Future work should explore richer statistical signals, diversified datasets, alternative join strategies, and more expressive predictive models.

*5.2 Future Work*

Although this study demonstrates that pre-join uniqueness can serve as a simple and effective signal for predicting changes in identifiability, several avenues remain open for future exploration. First, the model can be extended by incorporating additional pre-join statistical features—such as entropy, cardinality overlap, and distribution similarity—to improve predictive accuracy and capture more complex integration behaviours. Second, applying the method to larger, more diverse datasets and different types of joins (e.g., left, right, multi-key joins) may reveal patterns of identity leakage that do not emerge under simple inner-join conditions.

Another promising direction is developing a fully automated "privacy risk estimator" that integrates with organizational data pipelines, enabling real-time leakage prediction before data sharing or integration occurs. Finally, future research could explore how data harmonization, anonymization techniques, and synthetic data generation interact with uniqueness-based leakage prediction. These extensions would provide a broader understanding of identifiability dynamics and strengthen the practical value of pre-join leakage assessment.

## 6. COCNLUSION

This study investigated whether identity-leakage risk can be predicted *before* performing a dataset join operation. By using uniqueness as a pre-join statistical signal and training a supervised model based on Gradient Boosted Trees, the approach successfully estimated the direction of identifiability change without requiring an actual integration of sensitive data. Experimental results showed that the join decreased uniqueness in the combined dataset, and the model correctly predicted this direction using only the pre-join uniqueness of the two source datasets.

These findings demonstrate that organizations can evaluate potential leakage risks without exposing or merging their datasets, enabling safer integration planning. The work highlights the importance of understanding how dataset characteristics influence identifiability, and provides a lightweight approach for pre-join privacy assessment.

## 7. BIBLIOGRAPHY

[1] Richman, A. (2025, November 23). *Re-identification of anonymized data: What you need to know*. From k2view: https://www.k2view.com/blog/re-identification-of-anonymized-data

[2] Wong, W., Alomari, Z., Liu, Y., & Jora, L. (2024). Linkage Deanonymization Risks, Data-Matching and Privacy: A Case Study. *ResearchGate*, 6.

[3] Polonetsky, J. (2025, May 5). *The Curse of Dimensionality: De-identification Challenges in the Sharing of Highly Dimensional Datasets*. Retrieved December, 2025 from Future of Privacy Forum (fpf): https://fpf.org/blog/the-curse-of-dimensionality-de-identification-challenges-in-the-sharing-of-highly-dimensional-datasets/

[4] Sweeney, L. (2002). k-ANONYMITY: A MODEL FOR PROTECTING PRIVACY. *International Journal on Uncertainty, Fuzziness and Knowledge-based Systems, 5*(10), 557-570.